\documentclass[reprint,aps,pra,showpacs,floatfix]{revtex4-1}
\usepackage{graphics,epsfig,graphicx}
\usepackage{amsbsy}
\usepackage{amsmath}
\usepackage{bm}
\usepackage{amsfonts}
\usepackage{multirow}
\begin{document}

\title{Matching universal behavior with potential models}

\author{R. \'Alvarez-Rodr\'{\i}guez}
\affiliation{Escuela T\'ecnica Superior de Arquitectura, Universidad
Polit\'ecnica de Madrid, Avda. Juan Herrera 4, E-28040 Madrid, Spain}

\author{A. Deltuva}
\affiliation{Institute of Theoretical Physics and Astronomy, Vilnius University,
A. Go\u{s}tauto St. 12, LT-01108 Vilnius, Lithuania}

\author{M. Gattobigio}
\affiliation{Universit\'e de Nice-Sophia Antipolis, Institut Non-Lin\'eaire de
Nice,  CNRS, 1361 route des Lucioles, 06560 Valbonne, France }

\author{A. Kievsky} 
\affiliation{Istituto Nazionale di Fisica Nucleare, Largo Pontecorvo 3, 56100 Pisa, Italy}

 \begin{abstract}
Two-, three-, and four-boson systems are studied close to the unitary limit using
potential models constructed to reproduce the minimal information given by the 
two-body scattering length $a$ and the two-body binding energy or virtual state 
energy $E_2$. The particular path used to reach the unitary limit is given by varying the
potential strength. In this way the energy spectrum in the three- and four-boson systems
is computed. The lowest energy states show finite-range
effects absorbed in the construction of level functions
that can be used to study real systems. Higher energy levels are free from
finite-range effects, therefore the corresponding level
functions tend to the zero-range universal function. Using this property a zero-range
equation for the four-boson system is proposed and the four-boson universal
function is computed.
 \end{abstract}
\maketitle

\section{Introduction}

The Efimov effect has been predicted by V. Efimov in a series of 
papers~\cite{efimov1,efimov2} and experimentally confirmed more than 35
years after its prediction~\cite{kraemer2006}. At present there is 
an intense experimental activity~\cite{ferlaino2011,machtey2012,roy2013,dyke2013} 
aiming at understanding the behavior of few-body systems 
close to the unitary limit in which the two-body scattering length $a$ diverges.
Around this limit the systems show universal behavior; very different systems such
as atomic or nuclear systems present similar features. The study of these
characteristics is a very active field of research nowadays. 
From a theoretical point of view the behavior of a few-body 
system in the limit of large scattering length can be formulated in the
framework of the renormalization group using an Effective Field Theory 
(EFT)~\cite{bedaque99,bedaque2000} (for recent reviews see Refs.~\cite{report,frederico2011}).
Using this language, one of the main striking properties of three identical bosons
in the unitary limit is the discrete scaling invariance (DSI) shown by the
spectrum: an infinite series of bound states (trimers) appears distributed geometrically with 
accumulation point at zero energy. The ratio of binding energies for two consecutive states is
 $E^n_3/E^{n+1}_3=e^{2\pi/s_0}$, with the universal number $s_0\approx 1.00624$. 
Explicitly, the total angular momentum $L=0$ spectrum of three 
identical bosons in the zero-range limit can be
described by the Efimov radial law
\begin{subequations}
  \begin{eqnarray}
    \label{eq:energyzrA}
      E_3^n/(\hbar^2/m a^2) = \tan^2\xi , \\
      \kappa_*a = \text{e}^{(n-n^*)\pi/s_0} 
      \frac{\text{e}^{-\Delta_3(\xi)/2s_0}}{\cos\xi}\, .
    \label{eq:energyzrB}
  \end{eqnarray}
    \label{eq:energyzr}
\end{subequations}
The main ingredients in these equations are the universal function
$\Delta_3(\xi)$
and the binding momentum $\kappa_*$, called the three-body parameter,
 defining the energy $\hbar^2\kappa_*^2/m$ of level $n_*$
at the unitary limit, $m$ being the boson mass.
DSI manifests from the fact that
the function $\Delta_3(\xi)$ is the same for all $n$ levels.
Furthermore, the spectrum described by these equations is not
bounded from below, this characteristic is known as the Thomas
collapse~\cite{thomas1935}.

The determination of the universal function $\Delta_3(\xi)$ in the interval
$-\pi\le \xi \le -\pi/4$  limited by the two- and three-cluster continuum
can be obtained by solving
the Skorniakov-Ter-Martirosian (STM) equation or equivalently by using Effective
Field Theory (EFT) as reported in Ref.~\cite{bedaque99}. This equation describes
the $L=0$ state of a three-boson system in the zero-range limit. To avoid the Thomas
collapse it is a common practice to introduce a cutoff in the solution of the STM
equation and the universal function $\Delta_3(\xi)$ is computed looking at
the second or even third excited state~\cite{braaten2003} where the cutoff
effects are negligible. A parametrization of it can be found in 
Ref.~\cite{report}. In the same way potential models can be used to solve the
Schr\"odinger equation looking at the high part of the
spectrum, where finite-range effects are negligible.

The extension of the zero-range theory to four bosons has been discussed
in Refs.~\cite{platter2004,hammer2007,green2009,hadi2011}. The main conclusion
of these works is that the four-boson spectrum presents a two-level tree structure.
For each three-body level $E_3^n$ there are two four-body states (tetramers),
one deep ($m=0$) and one shallow ($m=1$), with binding energies $E_4^{n,m}$.
The universal ratios of these binding energies in the unitary
limit have been calculated in Ref.~\cite{deltuva2010} and they are
$E_4^{n,0}/E_3^n=4.611$ and $E_4^{n,1}/E_3^n=1.0023$.
In the present work we would like to analyze the complete
interval between the four-body continuum and the dimer-dimer thresholds in order 
to extend Eq.(\ref{eq:energyzr}) to the four-boson system and, if possible, to determine
the corresponding universal function.

The present study is done using potential models with variable strength.
In this way a path to reach the unitary limit is defined.
It has been recently shown that a two-parameter potential captures the
essential ingredients of the few-boson dynamics close to the unitary
limit~\cite{kievsky2015}. Accordingly, here we define two different potentials, a local
gaussian and a nonlocal gaussian used to solve the Schr\"odinger equation along the
path. For the lowest states finite-range effects are appreciable.
For this case it is possible to define level functions that
absorb those effects and can be used to estimate the spectrum of a real system
close to the unitary limit. For higher states finite-range effects can be
neglected and the solution tends to the zero-range limit therefore the level
functions tend to the universal zero-range function. In order to
illustrate the procedure we first analyse the two-body system and then
the analysis is extended to the three- and four-boson systems.

The paper is organized as follows. In Sec.II the particular
path used to reach the unitary limit is studied in the two-body system.
In Sec. III and IV the spectra of the three- and four-boson systems are 
computed using the two potential models, local and nonlocal. From the results a
zero-range equation for the four-boson system is proposed. The conclusion
and perspectives are given in the last section.

\section{Reaching the unitary limit with potential models} 

In the two-body system the $L=0$ spectrum within the zero-range theory can be defined 
through the relations
\begin{subequations}
  \begin{eqnarray}
    \label{eq:energy2bzrA}
      E_2=\frac{\hbar^2}{m a^2}, \\
      k\cot\delta_0=-\frac{1}{a}.
    \label{eq:energy2bzrB}
  \end{eqnarray}
    \label{eq:energy2bzr}
\end{subequations}
%where $a$ is the two-body scattering length. 
The first relation establishes that
there is a bound state ($a>0$) or a virtual state ($a<0$) with binding energy $E_2$ fixed 
by the scattering length $a$. At
positive energies, $E=\hbar^2k^2/m$, the $s$-wave phase-shift $\delta_0$
is determined by the scattering length as well. Accordingly,
the scattering length $a$ emerges as a control parameter in terms of which the
observables as the cross section or mean square radius can be computed. 
The zero-range theory describes the extreme situation in which the two
particles are always outside the interaction range. 
If the two-body quantum system, interacting through a short-range 
potential, has a shallow state, there is a big probability of finding 
the particles outside the interaction range.
In fact, when $E_2$ is very small or,
equivalently, $a_B = \hbar/\sqrt{mE_2}>>r_0$, with $r_0$ the interaction range, 
the two-body wave function has a very long tail and the two particles have a large 
probability of being at relative distances greater than $r_0$.
When a shallow state is present, the scattering length verifies
$a>>r_0$ too, and $a\approx a_B$, with Eq.(\ref{eq:energy2bzrA}) 
approximately fulfilled. The extension of Eq.(\ref{eq:energy2bzr})
in the case of finite-range interactions and at low energies is
\begin{subequations}
  \begin{eqnarray}
    \label{eq:energy2bfrA}
       E_2=\frac{\hbar^2}{m a_B^2} \hspace{1cm}, \\
       k\cot\delta_0=-\frac{1}{a}+\frac{1}{2}r_{\rm eff}k^2 \; ,
    \label{eq:energy2bfrB}
  \end{eqnarray}
    \label{eq:energy2bfr}
\end{subequations}
with $r_{\rm eff}$ being the effective range.
In the case of shallow states the second equation can be used to
relate the effective range to $a_B$,
\begin{equation}
\frac{1}{a_B} = \frac{1}{a}+\frac{r_{\rm eff}}{2a_B^2} \;\ ,
\end{equation}
from which we obtain the relation
\begin{equation}
r_{\rm eff}a=2r_Ba_B \;\; .
\label{eq:reff}
\end{equation}
with $r_B=a-a_B$. The above discussion stressed the fact that in 
the low energy limit or large scattering
length limit the dynamics of the two-body system depends 
on two parameters: the scattering length and the effective
range $r_{\rm eff}$ (or the length $r_B$).

The scaling limit is defined by $r_B\rightarrow 0$ whereas
in the unitary limit $1/a$ and $1/a_B \rightarrow 0$. 
In the first case, for each value of $a$ the two-body energy 
is determined by the zero-range condition $a=a_B$. When
$r_B\ne 0 $, the unitary limit can 
be reached by different paths determined by the functional relation $a_B=a_B(a)$. 
Having in mind that we intend to study the structure of few-boson systems, we
construct a two-parameter potential able to reproduce the 
minimal information given by one specific set of $a$ and $a_B$ values. Then the
potential strength can be varied in order to reach the unitary limit. 
We define local (given in coordinate space) and nonlocal (given in momentum space) 
gaussian potentials
\begin{eqnarray}
& V&^L_\lambda(r)=-\lambda V^L_0 e^{-r^2/r_0^2} \;\; ,
\label{eq:pot1} \\
& V&^{NL}_\lambda(k,k')=-\lambda V^{NL}_0 e^{-k^2/k_0^2} e^{-{k'}^2/k_0^2} \;\; ,
\label{eq:pot2}
\end{eqnarray}
with the strengths $V^L_0$, $V^{NL}_0$ and the ranges $r_0$, $k_0^{-1}$
determined to describe the particular value of $a$ and $a_B$ of a two-boson
system. The parameter $\lambda$ can be varied in order
to reach the unitary limit. With this procedure the functional relation
$a_B=a_B(a)$ is determined. It should be noted that
with the potentials defined above the lengths, momenta and energy scale with 
$r_0$, $k_0$ and $\hbar^2/mr_0^2$ (or $\hbar^2 k_0^2/m$), respectively. Accordingly,
the local gaussian defines a particular path to the unitary limit that
encompasses all local gaussians and the same for the nonlocal one. In particular,
the values of the effective range and strength at unitary are given in
table~\ref{tab:tab1} for the cases in which there are $n$ bound states
in the two-body system. Note that the nonlocal potential, being rank-one separable potential,
supports only one bound state at most.

\begin{table}[h]
\caption{Universal values of the effective range (in units of $r_0$ or $k_0^{-1}$) and potential 
strength (in units of $\hbar^2/mr_0^2$ or $\hbar^2 /k_0m$) for local and nonlocal
gaussians at the unitary limit for different bound state numbers  $n$.}
\begin{tabular}{c | l l | l l}
     & local            &                    &  nonlocal          &  \\
 $n$ & $r_{\rm eff}/r_0$    & $\lambda V^L_0mr_0^2 /\hbar^2$   &  $r_{\rm eff} k_0$     &
$\lambda V_0^{NL} m k_0/\hbar^2 $   \\
\hline
  0  &  1.43522         &   2.6840           &  3.19154     &  0.126987                  \\
  1  &  2.41303         &   17.7957          &                    &                      \\
  2  &  2.89034         &   45.5735          &                    &                      \\
  3  &  3.20006         &   85.9632          &                    &                      \\
\hline
\end{tabular}
\label{tab:tab1}
\end{table}

\begin{figure}[h]
\vspace{0.8cm}
    \begin{center}
      \includegraphics[width=\linewidth]{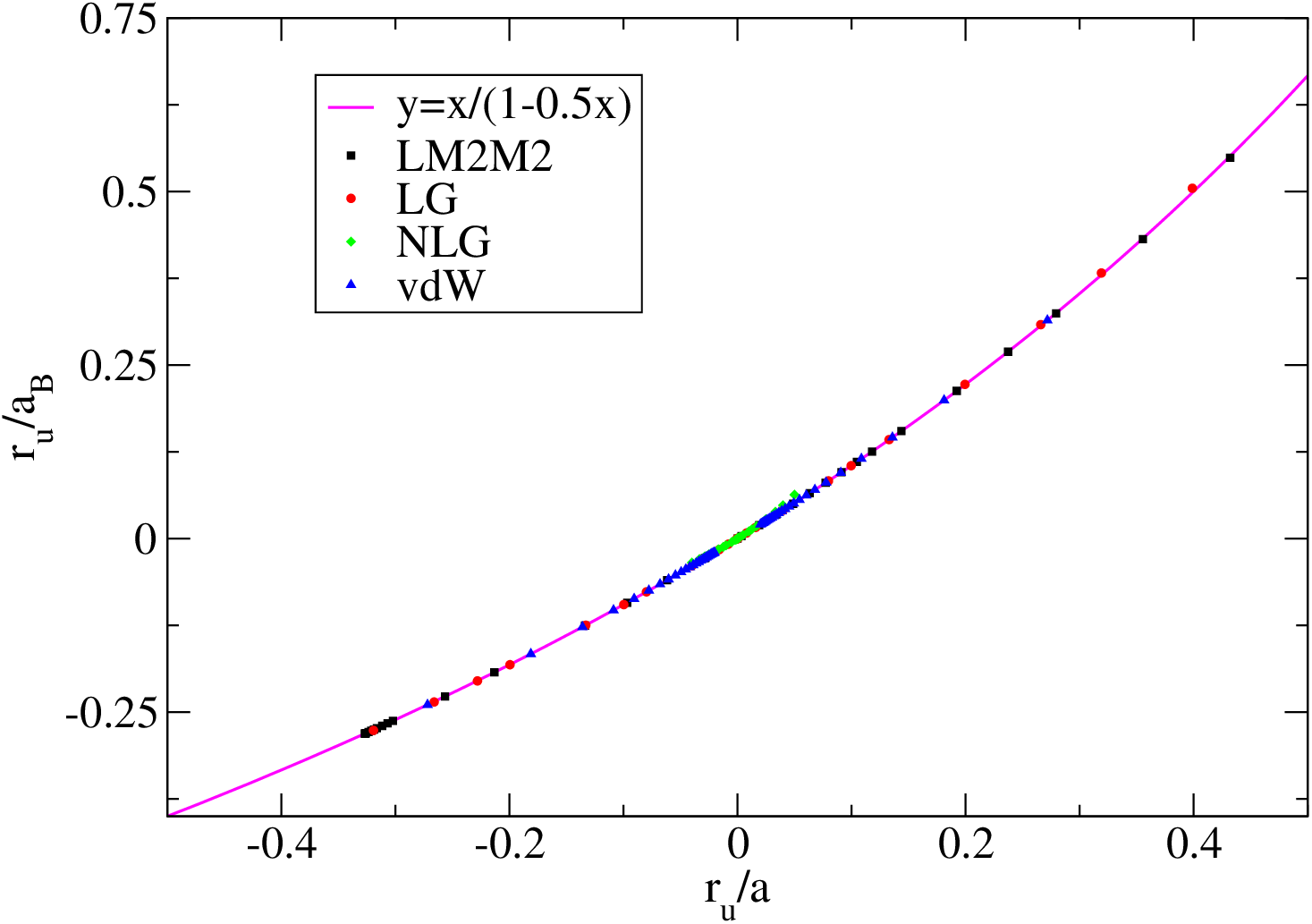}
    \end{center}
    \caption{(color online). The relation between the inverse of $a_B$ and
 $a$ (in units of $1/r_u$) for the different potential models. All the
computed values collapse on the curve $y=x/(1-0.5x)$. 
     }
\label{fig:fig2}
\end{figure}

In the following we consider the $n=0$ case, however the other cases can
be analyzed in a similar way. We are interested in the functional relation
$a_B(a)$ as $1/a\rightarrow 0$. 
Defining $r_u$ the value of the effective range at the unitary limit, from
Eq.(\ref{eq:reff}) we can define
\begin{equation}
\frac{r_{\rm eff}}{r_u}=\frac{2r_B}{r_u}-\frac{x}{2}\left(\frac{2r_B}{r_u}\right)^2
\label{eq:reffru}
\end{equation}
with $x=r_u/a$. Moreover, the quantity $2r_B/r_u$ can be expanded
around the unitary limit as
\begin{equation}
\frac{2r_B}{r_u}=1+{\cal A}x+{\cal B}x^2+\ldots .
\end{equation}
Inserting the above expansion in Eq.(\ref{eq:reffru}) the effective range
 expansion in terms of the inverse of the scattering length  becomes
\begin{equation}
\frac{r_{\rm eff}}{r_u}=1-{\cal C}x+{\cal D}x^2\ldots .
\label{eq:reffx}
\end{equation}
If the length $r_B$ were constant along the path, the coefficient ${\cal C}$
would be $0.5$ and ${\cal D}=0$ as well as higher terms. 
Differences from these values indicate a
non constant behavior of $r_B$. For example the study of the van der Waals (vdW)
potential  shows a quadratic relation between $r_{\rm eff}$ and
the inverse of the scattering length. Explicitly it results
(see Ref.~\cite{chin2010} and references therein)
\begin{equation}
 \frac{r_{\rm eff}}{r_u}=1-\frac{12\pi^2}{\Gamma(1/4)^4}x+\frac{72\pi^4}{\Gamma(1/4)^8}x^2
\end{equation}
with $\Gamma(1/4)^4/(6\pi^2)\approx 2.9179$. Therefore the coefficients in
Eq.(\ref{eq:reffx}) are ${\cal C}\approx 0.685$
and ${\cal D}\approx 0.235$. In the case of
the LG and NLG potentials the behavior is almost linear
(${\cal D}\approx 0$), with the coefficient ${\cal C}\approx 0.504$ and
$0.393$, respectively. 
The relation $a_B=a_B(a)$ can be analyzed starting from the following 
definition
\begin{equation}
  \frac{r_u}{a_B}=\frac{r_u}{a}\frac{1}{1-r_B/a} \,\, .
\end{equation}
Defining $x=r_u/a$ and $y=r_u/a_B$, and considering the
expansion of $2r_B/r_u$ given above, this relation, at first order in $1/a$, results
\begin{equation}
  y=\frac{x}{1-0.5x} \,\, .
\label{eq:unixy}
\end{equation}
In Fig.~\ref{fig:fig2} the quantity $r_u/a_B$ is shown for different potential
models around the unitary limit. The results of the different potential models
collapse in the curve given by
Eq.(\ref{eq:unixy}) showing that up to first order the relation
$a-a_B\approx 0.5r_u$ is well verified. In the study we have included
the LM2M2 Helium-Helium interaction from Aziz~\cite{aziz1991}.
We can conclude that Eq.(\ref{eq:unixy}) can be
seen as a universal relation describing the path to the unitary limit
fixed by the relation $a-a_B=\;constant$. Potentials with variable strength follow
this path with reasonable accuracy, however first order corrections could be of
the order of a few percent for the LG potential and up to $10\%$ in the case of the
NLG potential.

\section{Universal behavior in the three-body sector} 

The analysis of the two-boson system indicates that a two-parameter
potential as  local or non-local gaussian can be used
to study the low-energy dynamics around the unitary limit. We will extend
the analysis to the three- and four-boson systems.
The numerical results for the local gaussian are obtained solving the Schr\"odinger equation
in the coordinate space framework using the Hyperspherical Harmonic 
expansion~\cite{XXX1,XXX2,XXX3} while
the predictions for the non-local gaussian are obtained solving  Faddeev-Yakubovsky (FY)
\cite{yakubovsky:67}
or  Alt-Grassberger-Sandhas (AGS) equations \cite{grassberger:67}
using momentum-space methods from Refs.~\cite{deltuva2010,deltuva:11a}.

The $L=0$ spectrum of three identical bosons in the zero-range limit can be described
by the Efimov radial law given in Eq.(\ref{eq:energyzr}). In the case of a
system with a finite range interaction, this equation can be
seen as describing the asymptotic spectrum of the three-boson system 
close to the unitary limit. In fact, the solution of the Schr\"odinger equation 
can be used to determine the universal function $\Delta_3(\xi)$ 
looking at the excited states of the spectrum as, for these states, finite-range 
effects are negligible. In this context the description of few-body systems with
potential models close to the unitary limit can be seen as a particular regularization scheme.
Accordingly it is possible to modify the Efimov radial law as
(see Refs.~\cite{kievsky2015,gatto2014})
\begin{subequations}
  \begin{eqnarray}
    \label{eq:energyfr3A}
      E_3^n/E_2= \tan^2\xi, \\
      \kappa^n_3a_B = \frac{\text{e}^{-\widetilde\Delta^n_3(\xi)/2s_0}}{\cos\xi}\, ,
    \label{eq:energyfr3B}
  \end{eqnarray}
    \label{eq:energyfr3}
\end{subequations}
where $E_2=\hbar^2/m a_B^2$ is the dimer binding energy
for positive values of $a$ whereas for negative
values it is the two-body virtual state energy. Modifications at the three-body
level are introduced by the parameters $\kappa^n_3$ who absorb the scaling factor 
$e^{n\pi/s_0}$ defining the energy of level $n$ at the unitary limit,
$E^n_u=(\hbar^2/m)(\kappa^n_3)^2$. Furthermore the
finite-range character of the interaction slightly modifies the ratio
$\kappa^n_3/\kappa_3^{n+1}$ from its universal value of $\approx 22.7$. 
The main modification in the above equations is the introduction of
the level function $\widetilde\Delta^n_3(\xi)$. For the ground
state ($n=0$) it could be very different from the zero-range function 
$\Delta_3(\xi)$. As we will see below the differences are much reduced considering
the first excited state ($n=1$) and, starting from $n>1$, both functions almost coincide.
The level function can be calculated using the 
corresponding solutions of the Schr\"odinger equation as
\begin{equation}
\widetilde\Delta^n_3(\xi)=s_0\ln\left(\frac{E_3^n+E_2}{E^n_u}\right).
\end{equation}

\begin{figure}[h]
\vspace{0.8cm}
    \begin{center}
      \includegraphics[width=\linewidth]{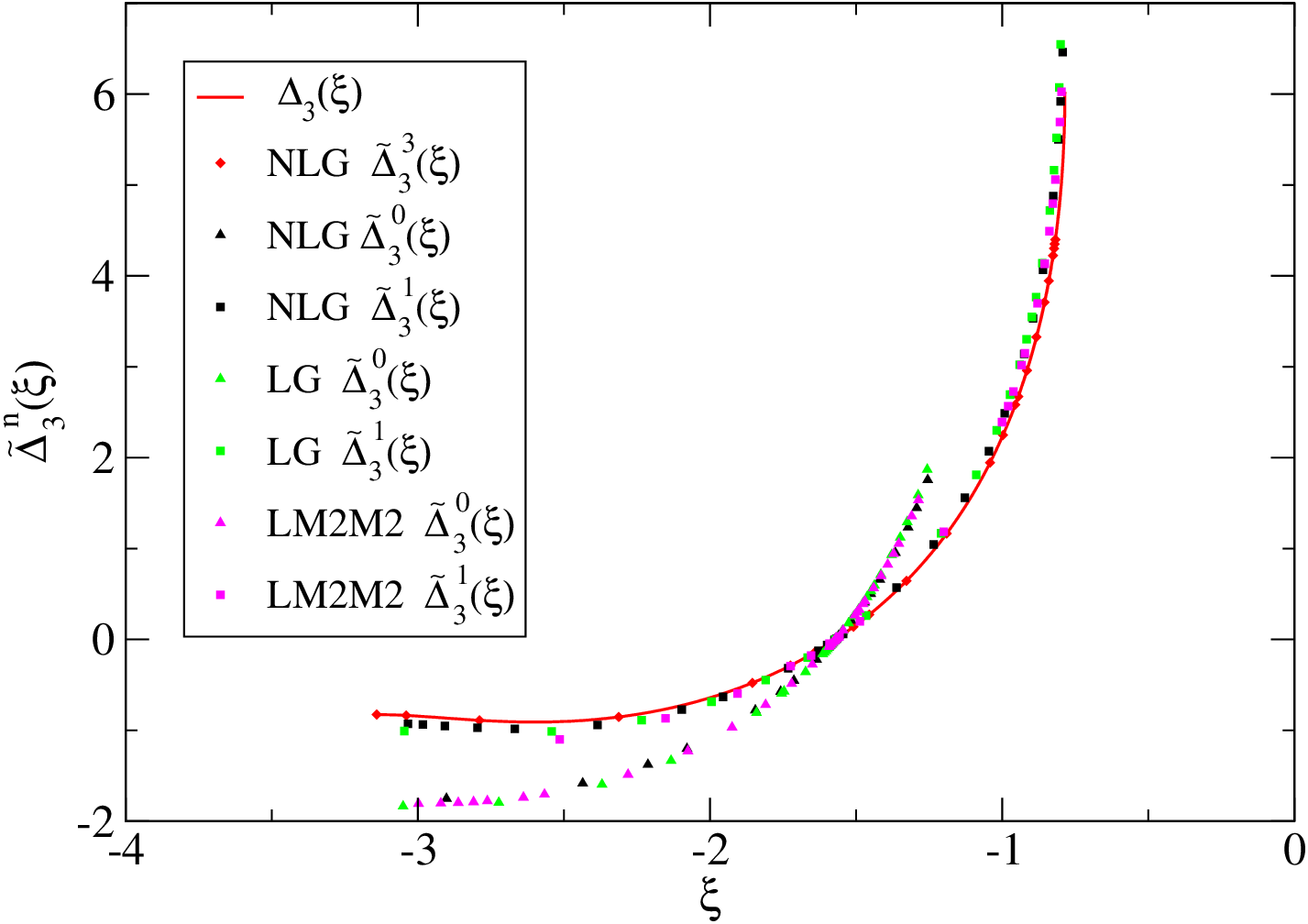}
    \end{center}
    \caption{(color online). The level functions $\widetilde\Delta^n_3(\xi)$
  for different levels and potentials.
 The zero-range universal function $\Delta_3(\xi)$ is shown as the red
solid curve.
     }
\label{fig:fig3}
\end{figure}

It should be noticed that for $n=0,1$ this function depends on the particular potential used
to calculate the spectrum or, in the case of the STM equation, the cutoff.
It depends also on the particular path selected to
reach the unitary limit, for example, potentials with variable strength as
discussed in the previous section. 
However, as shown in Ref.~\cite{kievsky2015}, following this particular path
different potentials do not produce too much spread in $\widetilde\Delta^0_3(\xi)$
and $\widetilde\Delta^1_3(\xi)$. Furthermore
the LG potential defines a unique gaussian function $\widetilde\Delta^n_3(\xi)$
for each level $n$ independent of the range, $r_0$, of the potential. In
particular, for the first two levels $n=0,1$, the binding momenta at the unitary limit 
are $\kappa_3^0=0.4874/r_0$ and $\kappa_3^1=0.0212/r_0$ 
and the ratio $\kappa_3^0/\kappa_3^1\approx 23.0$. These values have been
obtained with the potential acting only in $s$-waves, they are
slightly different when the LG potential
is taken to act in all waves (see for example Refs.~\cite{kievsky2013,kievsky2015}). 
Also the NLG potential acting in $s$-wave defines a unique nonlocal gaussian function
with the following universal ratios $\kappa_3^0/k_0=0.2127$,
$\kappa_3^1/k_0=0.009085$ and $\kappa_3^0/\kappa_3^1\approx 23.4$. 
From the above discussion the following picture emerges: a two-parameter
potential as the local or nonlocal gaussians can be used to construct 
level functions for each level $n$ of the three-boson system.
For $n=0,1$ these functions are different from the zero-range universal
function and are also different among themselves. For $n>1$ they 
converge to the zero-range universal function showing a universal behavior.
In order to  analyze this fact quantitatively, in Fig.~\ref{fig:fig3} the
level functions $\widetilde\Delta^0_3(\xi)$ and
$\widetilde\Delta^1_3(\xi)$ are shown for the case of the LG, NLG and LM2M2
potentials. The trend discussed above is visible in the figure, for the
ground state the level functions $\widetilde\Delta^0_3(\xi)$ spread
in a narrow band and are rotated with respect to the universal function 
$\Delta_3(\xi)$, given by the red solid curve~\cite{hammer2016}. In the case of the
first excited state the level functions $\widetilde\Delta^1_3(\xi)$
spread also in a very narrow band very close to $\Delta_3(\xi)$.
The level function $\widetilde\Delta^3_3(\xi)$
calculated using the NLG third excited state
(red diamonds) completely overlaps with the zero-range universal function.

The Efimov radial law is a one-parameter equation. The knowledge of the
universal function $\Delta_3(\xi)$ allows for a complete determination of the spectrum
after assigning a value to $\kappa_*$ (or to one of the energies $E_3^n$).
Eq.(\ref{eq:energyfr3}) applies mostly to $n=0,1$ and works slightly differently. 
In first place it is necessary to calculate the level functions for the ground 
and first excited state. This can be done using for example the LG or NLG potentials. 
At this point the equation is a one-parameter equation and the $n=0$
and $n=1$ spectrum can be completely determined from the knowledge of one energy.
One can argue that if we use Eq.(\ref{eq:energyfr3}) to describe a particular
system, the respective potential could be used to compute the
level function $\widetilde\Delta^0_3(\xi)$ or $\widetilde\Delta^1_3(\xi)$ 
by varying the strength in order to reach the unitary limit. However, as it is
shown in Fig.~\ref{fig:fig3}, close to the unitary limit the three-boson system has
universal behavior and therefore a two-parameter potential captures the essential 
ingredients of the dynamics absorbing finite range effects. As an example we
can use $\widetilde\Delta_0(\xi)$ computed using the LG or NLG potentials
to estimate the three-body parameter $\kappa^0_3$ of a system composed by three
$^4$He atoms. As given in Ref.~\cite{kievsky2015} the result using the LG potential
is $\kappa^0_3\approx 0.0438\;$a$_0^{-1}$ (a$_0=0.529177\ldots\;$ \AA\  is the
Bohr radius). Using the NLG we obtain
$\kappa^0_3\approx 0.0442\;$a$_0^{-1}$ whereas with the LM2M2 potential the
result is $0.0440\;$a$_0^{-1}$. This shows that the level functions
produce a description with the accuracy better than $1\%$ . The results for the three-body
parameter corresponding to the first excited state are
$\kappa^1_3\approx 0.0018\;$a$_0^{-1}$ and $\kappa^1_3\approx 0.0019\;$a$_0^{-1}$
for the LG and NLG respectively. To be compared to the LM2M2 value of
$\kappa^1_3=0.0019\;$a$_0^{-1}$. As expected finite range effects are
reduced in this level.
 
\section{Universal behavior in the four-body sector} 

\begin{figure}[h]
\vspace{1.2cm}
    \begin{center}
      \includegraphics[width=\linewidth]{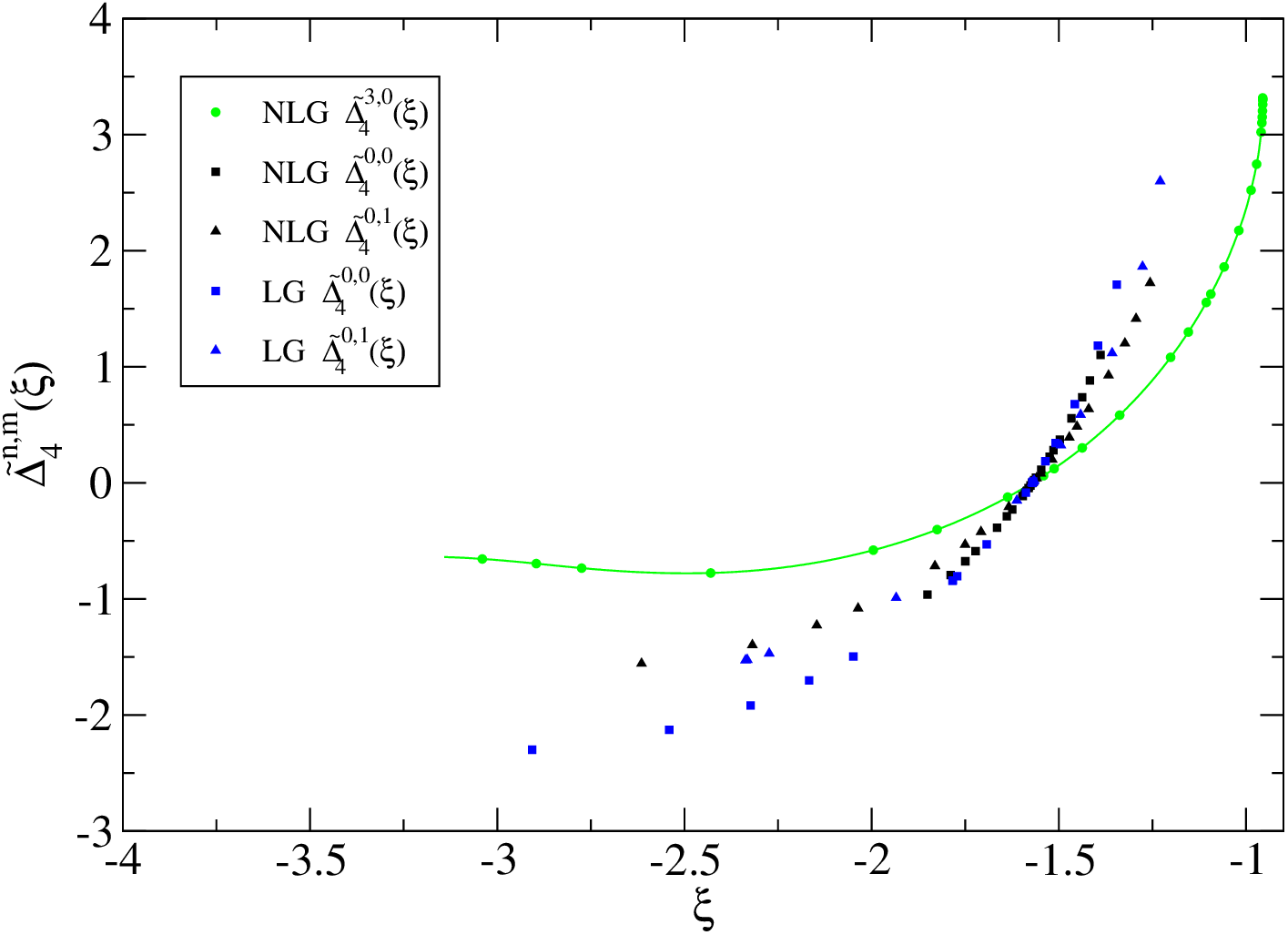}
    \end{center}
    \caption{(color online). The level functions $\widetilde\Delta^{n,m}_4(\xi)$
 for different levels and potentials.
The solid green line is an interpolation of the $\widetilde\Delta^{3,0}_4(\xi)$
values and should represent the universal function $\Delta_4^0(\xi)$.
     }
\label{fig:fig4}
\end{figure}

The previous analysis can be extended to the four-body case. In this case
a two-level structure, with energies $E_4^{n,0}$ and $E_4^{n,1}$, is attached 
to each $E_3^n$ level~\cite{platter2004,hammer2007,green2009,hadi2011}. 
As in the three-body case, the four-body system can be studied using potential
models, also in this case there is a modification of the universal ratios due to finite-range
effects (see for example Refs.~\cite{deltuva:11a,kievsky2014}).
Following the previous discussion, the equations describing the four-boson
spectrum can be written as
\begin{subequations}
  \begin{eqnarray}
    \label{eq:energyfrA4}
      E_4^{n,m}/E_2 = \tan^2\xi \;\; , \\
      \kappa^{n,m}_4 a_B = \frac{\text{e}^{-\widetilde\Delta^{n,m}_4(\xi)/2s_0}}{\cos\xi}\, .
    \label{eq:energyfrB4}
  \end{eqnarray}
    \label{eq:energyfr4}
\end{subequations}
with $m=0,1$ and $n$ identifying the corresponding three-body branch. The four-body
parameters $\kappa^{n,m}_4$ are related to the energy of the level at the
unitary limit, $E^{n,m}_u=\hbar^2(\kappa_4^{n,m})^2/m$. It should be stressed that only
the branch $n=0$, with energies $E_4^{0,0}$ and $E_4^{0,1}$,
corresponds to true bound states. The other states, corresponding to branches
with $n>0$, are above the trimer ground state threshold and therefore are 
unstable bound states (UBS)~\cite{deltuva:11a,kok}. 
In the above equation we have introduced the level
function $\widetilde\Delta^{n,m}_4(\xi)$ that governs the four-body spectrum
in levels $n,m$. They can be computed using potential models using the
following definition
\begin{equation}
\widetilde\Delta^{n,m}_4(\xi)=s_0\ln\left(\frac{E_4^{n,m}+E_2}{E^{n,m}_u}\right)
\; .
\end{equation}
For $n=0$ it could be very different from the universal function $\Delta^{m}_4(\xi)$
that governs the four-body spectrum in the zero-range limit. However, as $n$
increases finite-range effects become negligible and
$\widetilde\Delta^{n,m}_4(\xi)$ should tend to that function. 
The results are shown in Fig.~\ref{fig:fig4} where the function
$\widetilde\Delta^{n,m}_4(\xi)$ has been calculated using the LG and NLG
potentials for the ground state, $n=0,m=0$ level (squares) and first excited
state, $n=0,m=1$ level (triangles). For the sake of comparison 
the function corresponding to the $n=3,m=0$ level (circles) and calculated
using the NLG is also shown. From the figure we can see that the results
of both potentials for the $n=0,m=0$ and $n=0,m=1$ levels are close to each other. 
However there is a difference between the functions with different $m$ values,
more pronounced for $\xi < -\pi/2$.

Eq.(\ref{eq:energyfr4}) works very similar to the three-body case discussed
before. The knowledge of $\widetilde\Delta^{n,m}_4(\xi)$ makes this
equation a one parameter equation. We would like to stress that
$\widetilde\Delta^{n,m}_4(\xi)$ can be computed using a LG or a NLG potential
and then used to determine the spectrum of a real system around the unitary
limit. As an example we discuss the spectrum of four He atoms. A very complete
discussion of this system has been given in Refs.\cite{hiyama2012,hiyama2014}
using realistic potentials. The binding energies of the tetramers using the
LM2M2 interaction are $E^{0,0}_4=559.22\;$mK and $E^{0.1}_4=127.42\;$mK.
The dimer energy is $E_2=1.3094\;$mK and, therefore the angles are
$\xi=-1.52244$ and $-1.46977$ respectively. Using the NLG
$\widetilde\Delta^{n,m}_4(\xi)$ function we estimate 
$E_u^{0,0}\approx 0.443\;$K and 
$E_u^{0,1}\approx 0.0865\;$K in a very good agreement with the
quoted values for the LM2M2 potential of 
$E_u^{0,0}\approx 0.4449\;$K and $E_u^{0,1}\approx 0.0870\;$K
given in Ref.~\cite{hiyama2014}. We can conclude that the estimates
obtained using the level functions are given with an accuracy
well below $1\%$.

\begin{figure}[h]
\vspace{1.2cm}
    \begin{center}
      \includegraphics[width=\linewidth]{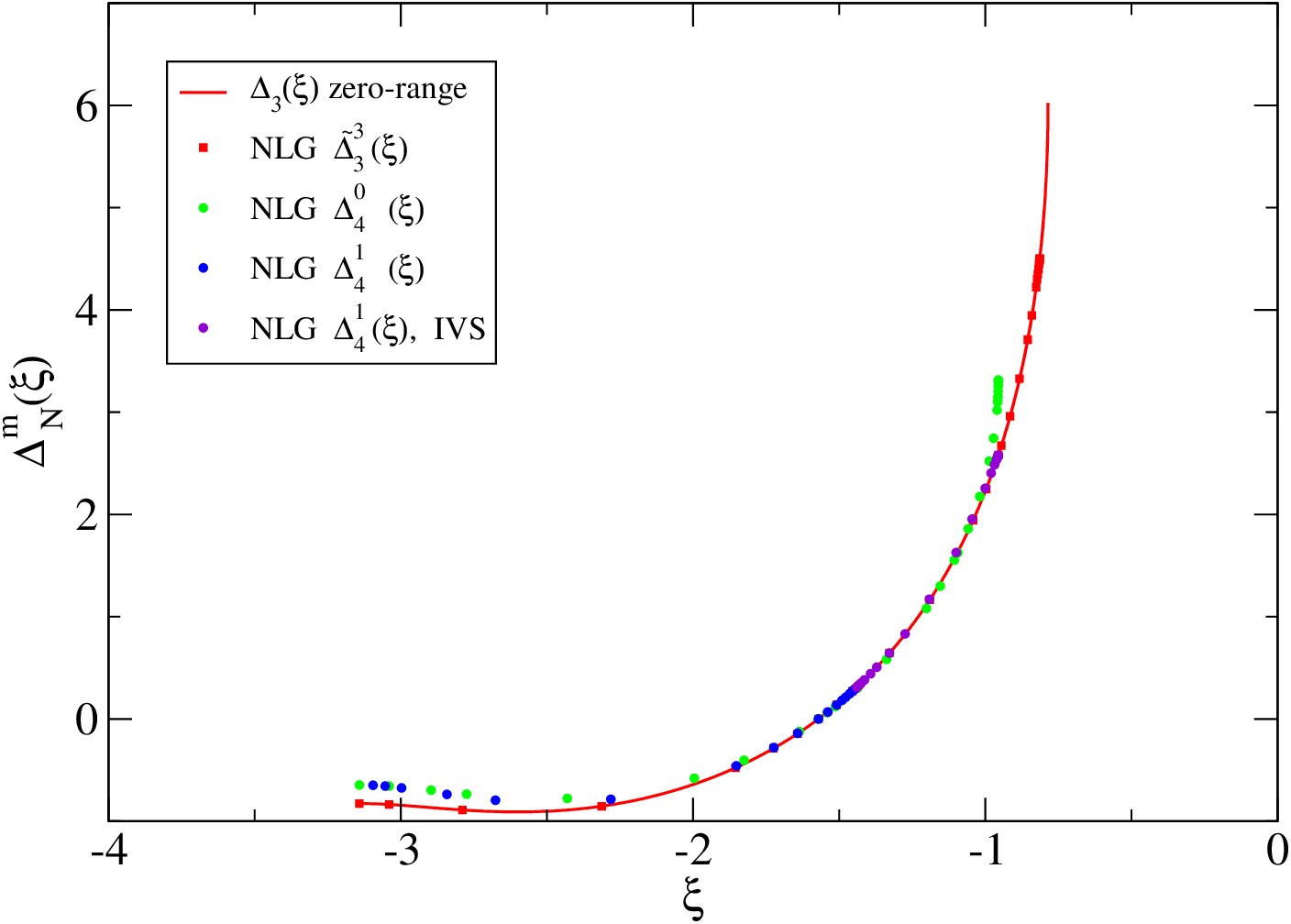}
    \end{center}
    \caption{(color online). The universal functions $\Delta^0_4(\xi)$ (green
 solid points) and $\Delta^1_4(\xi)$ (blue solid points and violet solid
 points in the IVS region). For the sake
 of comparison, the three-boson
 universal function $\Delta_3(\xi)$ and level function $\Delta^3_3(\xi)$
 are also shown.
     }
\label{fig:fig5}
\end{figure}

In Fig.~\ref{fig:fig4} the level function
$\widetilde\Delta^{3,0}_4(\xi)$, calculated using the NLG potential, is shown
(circles). For this level finite-range effects can be neglected and, therefore,
we can consider this function a good representation of the zero-range four-body 
universal function in the level $m=0$. For a generic level $m$ we define this function 
$\Delta_4^m(\xi)$. It does not
depend on the three-body branch $n$, as DSI, with the geometrical factor
$e^{\pi/s_0}\approx 22.7$, has been already verified among these
branches~\cite{green2009,deltuva2010,deltuva2012}. In the following we
study its dependence on the levels
$m$ and its relation with the three-boson universal function
$\Delta_3(\xi)$. To this aim the level functions
$\widetilde\Delta^{3,0}_4(\xi)$ and $\widetilde\Delta^{3,1}_4(\xi)$, 
calculated using the NLG potential, are shown in Fig.~\ref{fig:fig5} together
with the zero-range universal function $\Delta_3(\xi)$ and the level
function $\widetilde\Delta^3_3(\xi)$. 
We consider $\widetilde\Delta^3_3(\xi)$ a representation of the zero-range 
universal function $\Delta_3(\xi)$ and we consider $\widetilde\Delta^{3,m}_4(\xi)$
a representation of $\Delta_4^m(\xi)$. For the $m=1$ case the range of $\xi$
values in which this level results an inelastic virtual state (IVS) is explicitly
shown. From the figure we can see that 
$\Delta_3(\xi)$ and $\Delta_4^m(\xi)$ are very close to each other around the
unitary limit. As the functions approach the different thresholds differences
appear. In the case of the threshold at $\xi=-\pi$ we have  
$\Delta_4^0(-\pi)=\Delta_4^1(-\pi)=-0.645$, appreciably different from
$\Delta_3(\xi)=-0.8266$. Defining $a^{n,m}_{4,-}$, the two-body scattering length
at which the four-boson system disappears into the four-body continuum, the
first relation establishes that
\begin{equation}
\kappa_4^{n,m}a^{n,m}_{4,-}=1.378
\end{equation}
is a universal number. Defining $a^n_{3,-}$ to be the two-body scattering length
at which the trimer disappears into the three-body continuum, the
second relation results
\begin{equation}
\kappa_3^na^n_{3,-}=1.508,
\end{equation}
confirming the highly accurate value given in Ref.~\cite{aminus}. 
Within the zero-range theory, 
the set of values at which the different branches disappear into the three- 
and four-body continumm can be determined from the above relations 
using the universal ratios~\cite{deltuva2012}
$a^{n+1}_{3,-}/a^n_{3,-}=22.694$, $a^{n,0}_{4,-}/a^n_{3,-}=0.4254$,
$a^{n,1}_{4,-}/a^n_{3,-}=0.9125$.

The dimer-dimer threshold is defined by $\tan\xi_c=-\sqrt{2}$ corresponding to
$\xi_c=-0.9553$. In Fig.~\ref{fig:fig5} we can observe that around this value
$\Delta^0_4(\xi)$ differs from $\Delta^1_4(\xi)$ and $\Delta_3(\xi)$, with these
last two functions close to each other. It should be noticed that in this region
$\Delta^1_4(\xi)$ is obtained from the energy of the IVS
except for the vicinity of $\xi_c$ where the shallow tetramer again becomes UBS 
before decaying through the dimer-dimer threshold. The values of
the universal functions at the critical value $\xi_c$ can be calculated using the
universal ratios of Ref.~\cite{deltuva:11b}: $a^{n,0}_{dd}/a^n_{dd}=0.3235$,
$a^{n,1}_{dd}/a^n_{dd}=0.99947$, $a^n_{dd}/a^n_d=6.789$ 
and the relation $\kappa^n_3 a_n^d=0.0707645$, with $a^{n,m}_{dd}$, $a^n_{dd}$
and $a_n^d$ the scattering lengths at which the level $n,m$ of the tetramer intersects 
the dimer-dimer threshold and at which the level $n$ of the trimer intersects the 
dimer-dimer and dimer thresholds, respectively.
Using these ratios in Eq.(\ref{eq:energyfr4}) it results
$\Delta^0_4(\xi_c)=3.316$ and $\Delta^1_4(\xi_c)=2.580$, respectively,
in complete agreement with the computed values shown in
Fig.~\ref{fig:fig5}. We can conclude that the small differences between
$\Delta^0_4(\xi)$ and $\Delta^1_4(\xi)$ around the
critical value are due to threshold effects.

From the above discussion we propose the following zero-range equation for 
the $L=0$ spectrum of four equal bosons

\begin{subequations}
  \begin{eqnarray}
    \label{eq:energyzrA4}
      E_4^{n,m}/(\hbar^2/m a^2) = \tan^2\xi \\
      \kappa_*^ma = \text{e}^{(n-n^*)\pi/s_0} 
      \frac{\text{e}^{-\Delta_4^m(\xi)/2s_0}}{\cos\xi}\, ,
    \label{eq:energyzrB4}
  \end{eqnarray}
    \label{eq:energyzr4}
\end{subequations}
with $\kappa_*^m$ the binding momentum of the level $n_*,m$ at the unitary
limit verifying the following universal ratio $\kappa_*^0/\kappa_*^1=2.1449$ 
and $\Delta_4^m(\xi)$ the universal four-boson universal function given in
Fig.~\ref{fig:fig5}. This equation extends the
Efimov radial law for three bosons to the four-boson system.

\section{Conclusions} 

In the present work we have analysed two-, three- and four-boson systems
close to the unitary limit. To this aim we have solved the Schr\"odinger
equation (or FY and AGS equations) using potential models with variable strength constructed to reproduce 
the minimal information given by the two-body scattering length $a$ and the 
two-body binding energy or virtual state energy $E_2$. It has been shown that 
a two-parameter interaction as a gaussian can capture the main ingredients
of the dynamics in this region. 
Moreover these type of potentials define
level functions independent of the range used to compute it. This
property can be used to construct level functions of general validity that
can be used to predict some characteristics of real systems along the
particular path used to reach the unitary limit.
The level functions $\widetilde\Delta_3^n$ and $\widetilde\Delta_4^{n,m}$ 
have two properties: when they are used in the lowest branches, $n=0,1$,
they absorb finite-range effects. This portion of the spectrum does not
show a perfect DSI since finite-range effects are visible. 
So the interest here is to use the level functions to describe the dynamics
of real systems close to the unitary limit. For example potentials with variable strength
describe with reasonable accuracy the variation of the interatomic 
potential using broad Feshbach resonances in ultra cold atomic traps.

The second property is given by the description of the asymptotic part
of the spectrum. For levels with $n>1$ the spectrum calculated with the
potentials starts to show DSI and coincides with the spectrum in the
zero-range limit. Accordingly, the level functions for $n>1$ do not
depend any more on the level number $n$ and on the particular potential used to
compute it 
as well as the path selected to reach the unitary limit. They are good representations
of the universal functions $\Delta_3$ and $\Delta_4^m$. This property
has been used here to propose Eq.(\ref{eq:energyzr4}) as an extension of the
Efimov radial law for four bosons. 

In the present work we have studied a particular path to reach the unitary
limit based on a single channel potential with variable strength. Other possibilities 
could be for example the study of coupled channel interactions as in molecular 
systems. 
In this way different level functions can be constructed allowing to
a systematic study of finite-range effects.
Other improvements of the present work could be the study
of the spectrum as the number of bosons increases. Preliminary results
along this line have been obtained~\cite{kievsky2014}. Finally we would like
to mention the recent study of Efimov physics in the three-body system having spin-isospin
degrees of freedom~\cite{kievsky2016}. The extension to the four-body system is
under way.

\begin{acknowledgements}
This work was partly supported by Ministerio de Econom\'{i}a y Competitividad
(Spain) under contracts MTM2015-63914-P and FPA2015-65035-P. Part of the
calculations of this work were performed in the high capacity cluster for 
Physics, funded in part by Universidad Complutense de Madrid and in part 
with Feder funds as a contribution to the Campus of International 
Excellence of Moncloa, CEI Moncloa. R.A.R thanks Ministerio de
Educaci\'on, Cultura y Deporte (Spain) for the “Jos\'e Castillejo” fellowship 
in the framework of Plan Estatal de Investigaci\'on Cient\'{i}fica y T\'ecnica y 
de Innovaci\'on 2013-2016.
\end{acknowledgements}

\end{document}